\documentclass{article}
\usepackage{dcase2023,amsmath,graphicx,url,times,booktabs, tabularx, color}
\usepackage{balance}

\def \am[#1]{\textcolor{blue}{AM: #1}}
\title{EVALUATING CLASSIFICATION SYSTEMS AGAINST SOFT LABELS WITH FUZZY PRECISION AND RECALL}

\name{Manu Harju, Annamaria Mesaros\thanks{This work was supported by Academy of Finland grant 332063 ``Teaching machines to listen".}}

\address{Computing Sciences, Tampere University, Tampere, Finland}

\begin{document}

\ninept
\maketitle

\begin{sloppy}

\begin{abstract}

Classification systems are normally trained by minimizing the cross-entropy between system outputs and reference labels, which makes the Kullback-Leibler divergence a natural choice for measuring how closely the system can follow the data. Non-binary references can arise from various sources, and it is often beneficial to use the soft labels for training instead of the binarized data. In addition to the cross-entropy based measures, precision and recall provide another perspective for measuring the performance of a classification system. However, the existing definitions for precision and recall require binary reference labels, and binarizing the data can cause erroneous interpretations and loss of information about the underlying data distributions. We present a novel method to calculate precision, recall and F-score without quantizing the data. The proposed metrics are based on fuzzy theory and extend the well established metrics, as the definitions coincide when used with binary labels. To understand the behavior of the proposed metrics we show numerical example cases and an evaluation of different sound event detection models trained on real data with soft labels.

\end{abstract}


\section{Introduction}
\label{sec:intro}

The target labels in classification tasks are usually presented in one-hot or multi-hot encoded form, indicating target classes
being either present or absent. This leads naturally to using binary representations for the data; however,  nothing prevents us for using non-binary values. The terms hard and soft labels are often used to make the distinction between the binary and non-binary cases. Soft labels can be derived from binary data e.g. by using label smoothing \cite{szegedy:rethinking} or data augmentation with mixup \cite{zhang:mixup}. Furthermore, non-binary values can be used to represent uncertainty in the original data. Using soft labels to present the annotators confidence can improve the model performance \cite{peterson:human, nguyen:learning} and help with ambiguous classes \cite{grossmann:beyond}. 

One particular case is in the DCASE 2023 Challenge, where the organizers provide soft labels in the sound event detection (SED) task.  The data was collected by splitting approximately 3-minute long recordings into 10-second clips with a one-second stride, and the clips were annotated on Amazon Mechanical Turk for temporally weak labels. Each clip was annotated by five different annotator, and therefore a single one-second segment in the middle part in the longer files got annotation from 50 workers. The collected labels were used to estimate competence values for the individual annotators, which were then used to compute weighted averages of the annotator opinions, with the resulting numbers designated as the \textit{soft labels} for one-second segments of audio \cite{martinmorato:strong}.

The challenge task is about training a sound event detection system using the soft labels, to investigate if leveraging information from the soft labels is beneficial for the acoustic models. However, the evaluation is done using hard labels and hard metrics. Converting soft labels into binary requires choosing a threshold value, and finding a good one is not a trivial task. The most straightforward way is to use 0.5 as the threshold, and this is also how the reference data for the challenge is binarized. However, as a consequence, six event classes out of 17 are left out from the evaluation, as there are not enough segments with a soft label value above the threshold.

Converting the soft reference labels into binary means discarding part of the information about the data. On the other hand, using soft labels for training was shown to improve accuracy and yield more structure for the more ambiguous classes in the learned feature space \cite{grossmann:beyond}. Furthermore, training models with soft labels was shown to improve the performance on out-of-sample data and to make the models more robust against adversarial attacks \cite{peterson:human}. 

In \cite{martinmorato:training}, soft labels have been used for training a SED, and the Kullback-Leibler divergences in the results indicate that the system output is closer to the reference labels than for the case when the same system is trained using the hard labels. Furthermore, using the midpoint of the value range as the threshold for each class independently improves the performance on less represented classes. Nevertheless, if the system performance is measured from binary references, we are missing out the uncertainty levels in the data. In an extreme case such harsh quantization renders the data unusable by converting all the labels into the same value. 

In this paper we propose a novel method to calculate precision and recall from softly-labeled data. The proposed metrics coincide with the hard metrics when used with binary labels, and thus can be seen as an extension of the hard precision and recall for handling soft references and system outputs. Using the soft labels directly allows measuring the system's ability to track the reference data, taking the metrics closer to the cross-entropy based ones. However, the soft precision and recall also provide similar information about the bias of the classifier as the hard metrics. 

Measuring precision and recall requires counting set sizes for the predicted outputs and the references. For example, in multilabel classification, for a single input sample we may have a set of predicted labels and a set of reference labels. Previous work for soft metrics involve the idea of the set elements having soft boundaries for several different scenarios. Such elements may overlap with each other, making measuring the set size more difficult. The work in \cite{franti:soft} presents a number of examples, including comparing sets of words and audio segmentation. In the word set comparison, syntactic and semantic similarities are measured between words to determine the overlap and \emph{soft cardinality} of the sets. In the segmentation example the onset and offset times are from continuous scale but the segments themselves are hardly labeled. In \cite{ziolko:fuzzy}, audio segments are converted into fuzzy sets with soft boundaries and fuzzy set theory is used to compute fuzzy precision and recall. However, in all these examples, the elements activities in the prediction and reference sets are binary. In contrast, the elements of interest in our case are well defined audio segments with clear boundaries, but the intensities of the predictions and reference data, i.e. the soft labels, come from the unit interval. 

This paper is organized as follows: Section \ref{sec:method} introduces the theoretical definitions of the soft precision and recall and a few numerical examples for understanding the behavior of the metrics in various scenarios. Section \ref{sec:results} shows a comparison of hard and soft metrics using the DCASE 2023 Challenge SED with soft labels task data, and the method and results are discussed in Section \ref{sec:discussion}. Finally, Section \ref{sec:conclusions} presents conclusions and future work.

\section{Soft precision and recall}
\label{sec:method}
\subsection{Definitions}
Let $X$ be the set of all elements to be classified, and $L \subseteq X$, $G \subseteq X$ be the sets of predicted labels and ground truth references, respectively. In the binary case, the sets $L$ and $G$ contain the positively labeled elements. Precision and recall can be defined using set theoretical notation \cite{franti:soft} with:

\begin{eqnarray}
    \mathrm{Precision} & = & \frac{|L \cap G|}{|L|}, \nonumber \\
    \mathrm{Recall} & = & \frac{|L \cap G|}{|G|}, \\
    \mathrm{F_1-score} & = & 2 \frac{|L \cap G|}{|L| + |G|}. \nonumber
\end{eqnarray}

A fuzzy set is defined by its membership function $\mu: X \rightarrow [0, 1]$ mapping each element of $X$ to its grade of membership. We interpret $L$ and $G$ as fuzzy sets through the soft labels: given the reference label value $y_i$ and prediction $\hat y_i$ for the $i$'th item $x_i$, the membership function values are:
\begin{equation}
\begin{array}{rcl}
\mu_L(x_i) & = & \hat y_i, \\
\mu_G(x_i) & = & y_i.
\end{array}
\end{equation}

We use the standard definition of intersection for fuzzy sets, namely the minimum of the membership function values. Finally, the fuzzy set cardinality is just the sum of the membership values, resulting in the following definitions for the soft precision, recall, and $F_1$-score
\begin{eqnarray}
    \mathrm{Precision} & = & \frac{\sum_i \min(\hat y_i, y_i)}{\sum_i \hat y_i}, \nonumber \\
    \mathrm{Recall} & = & \frac{\sum_i \min(\hat y_i, y_i)}{\sum_i y_i}, \\
    \mathrm{F_1-score} & = & 2 \frac{\sum_i \min(\hat y_i, y_i)}{\sum_i (\hat y_i + y_i)}. \nonumber
\end{eqnarray}
It is possible to use some other T-norm for the fuzzy intersection, but the advantage of using minimum is its idempotency, i.e. $\min(x, x) = x$. From idempotency it follows that if $\hat y_i = y_i$ for any value in the unit interval, then also the precision and recall for that item will be equal to 1.

\subsection{Examples}
\label{sec:examples}

\begin{table*}
\begin{center}
\vspace{-8pt}
\begin{tabular}{m{41mm}|ccc|ccc|c}
 & \multicolumn{3}{c|}{Hard} & \multicolumn{3}{c|}{Soft} & \\
 
  \centering Data points & P & R & F & P & R & F & KLD \\
    
    \hline    
    \includegraphics[width=40.0mm]{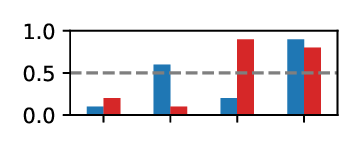} 
    & 50.0 & 50.0 & 50.0 & 66.7 & 60.0 & 63.2 & 0.446 \\

    \includegraphics[width=40.0mm]{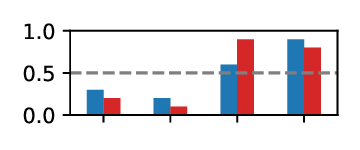}
    & 100.0 & 100.0 & 100.0 & 85.0 & 85.0 & 85.0 & 0.083 \\

    \includegraphics[width=40.0mm]{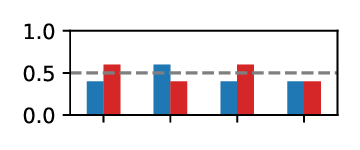}
    & 0.0 & 0.0 & 0.0 & 88.9 & 80.0 & 84.2 & 0.061
    
\end{tabular}
\caption{Simple example cases. Blue is predicted value, red is the reference label value. The dashed grey line is the threshold value at 0.5. The predictions are closest to the reference in the third case, even though the classical hard metrics indicate the opposite.}
\label{tab:examples}
\end{center}
\vspace{-10pt}
\end{table*}

\begin{figure}
    \centering
    \includegraphics[width=65.0mm]{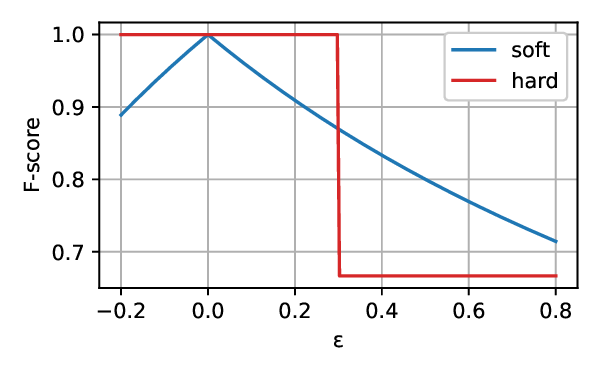}
    \caption{F-scores for a two-point example, comparing a 0.2 reference with a $0.2 + \varepsilon$ prediction. }
    \label{fig:param}
    \vspace{-10pt}
\end{figure}

Table \ref{tab:examples} shows three small example cases. For each case, four data points are illustrated as a bar plot, where the blue bars represent predictions and the red bars correspond to reference labels. The dashed grey line denotes the threshold value of 0.5. All three examples contain 50 \% of positive references in the binarized case. The table presents for each case the hard and soft metrics, and the last column shows Kullback-Leibler divergence for a measure of difference between the predictions and references. 

For the first two examples the reference labels are the same and they correspond to a more clear annotation of categories. In the first case, half of the predicted positives are correct, and half of the positive references are found;  in the second case, both positive references are found. For these two examples, the values for the hard metrics are as expected. However, the third example represents a more ambiguous case, where the labels are close to the 0.5 threshold value. In such a case the label quantization may turn the prediction and reference into complete opposites, resulting in very low scores despite the small difference between the system outputs and the reference labels. Furthermore, according to the KL-divergence the predictions are closest to the references in the last example, even though the hard metrics indicate the opposite.

To get a better insight about the behavior of the metric, we construct an example of two data points. The first data point is a true positive with value 0.8 for both prediction and reference. For the second point we set the reference label to 0.2 and use $0.2 + \varepsilon$ for our prediction. The effect of perturbing the prediction around the ground truth can be seen in Figure \ref{fig:param}. The figure shows that the soft F-score has a sharp peak at the correct value. For $\varepsilon > 0.3$ the hard metric interprets the label as incorrect, while the soft metrics still finds some common mass between the prediction and reference label, resulting in higher soft F-score than the hard one.

\section{Experimental results}
\label{sec:results}
\subsection{System evaluations}
\label{sec:systems}

We use the baseline system from the DCASE 2023 Challenge task 4B and two different modifications to explore the behavior of the proposed metrics. The baseline system is a standard CRNN with three convolutional blocks and one bidirectional GRU layer of size 32. Each convolutional block consists of a single 2D convolution with 128 filters and 3x3 kernels followed by batch normalization, rectified linear unit, max pooling, and dropout layers. The second system, denoted TinyCRNN, has two convolutional blocks, the number of filters is reduced to 32, and the GRU size is halved in comparison to the baseline. The third system, denoted DWSNet, uses a similar GRU layer and the same number of blocks as the baseline, but each block consists of four depthwise separable convolutions. The number of parameters for the baseline, TinyCRNN, and DWSNet models are 380k, 28k, and 215k, respectively. The models are trained with soft labels and regression setup using the development dataset and code provided with the baseline model.

In addition to the trained systems, we compute scores for three data-driven sampling methods for generating a randomized system output based on the training data. In the first case we fit beta distributions for each class in the training data and produce the system output by sampling from each distribution individually. The second sampled output is generated in a similar manner to the first one, but the fitted distributions are shuffled so that the classes are assigned a randomly chosen distributions of other classes. The original data comes from five different scenes with partially overlapping sets of classes, and the labels for absent classes are set to zero. As a result, the fitted distributions are biased towards zero, and for a single segment the method can output labels that are never seen co-occurring in the data. The third output is generated by sampling individual time steps from the training data, i.e. taking all outputs for all the classes at the same time.
 
For an extensive analysis, we use two different hard label evaluation methods that are supplied with the task baseline code. In the first case the system outputs are binarized using 0.5 as the threshold value. In the second case the system output values are automatically class-wise thresholded for the best macro F-score \cite{ebbers:threshold}. We call this method \textit{optimal threshold} (OT). The soft metrics are calculated over the set of 11 classes included in the challenge evaluation for a direct comparison with the above-mentioned metrics, and also for all the 17 classes that are present in the training data.

\begin{table*}
\centering
\vspace{-5pt}
\begin{tabular}{l|cc|cc|cc|cc|cc|cc}

& \multicolumn{2}{c|}{Betas}
& \multicolumn{2}{c|}{Shuffled betas}
& \multicolumn{2}{c|}{Data sampling} 
& \multicolumn{2}{c|}{Baseline} 
& \multicolumn{2}{c|}{TinyCRNN}
& \multicolumn{2}{c}{DWSNet}
\\
 & $F_{m}$ & $F_{M}$ 
 & $F_{m}$ & $F_{M}$ 
 & $F_{m}$ & $F_{M}$ 
 & $F_{m}$ & $F_{M}$ 
 & $F_{m}$ & $F_{M}$ 
 & $F_{m}$ & $F_{M}$ 

\\
\hline      
& & & & & & & & & & & &
\\

Hard         & 23.2 & 9.1  &  7.2 &  4.6 & 30.9 & 10.9 & 70.4 & 35.0 & 64.8 & 26.2 & 72.8 & 35.7 \\
Hard / OT    & 27.4 & 18.0 & 26.4 & 18.1 & 25.0 & 17.8 & 51.7 & 41.5 & 49.2 & 39.7 & 60.2 & 43.6 \\
Soft         & 35.5 & 23.0 & 17.1 & 14.1 & 39.0 & 25.9 & 72.6 & 65.5 & 69.4 & 60.2 & 77.0 & 70.3 \\
Soft (all classes)   & 33.3 & 18.6 & 15.7 & 12.6 & 36.8 & 21.1 & 70.4 & 58.8 & 66.8 & 53.2 & 75.1 & 64.5 

\\
& & & & & & & & & & & & 
\\
\hline

& & & & & & & & & & & &
\\

KLD
& \multicolumn{2}{c|}{0.746} 
& \multicolumn{2}{c|}{1.138} 
& \multicolumn{2}{c|}{0.074} 
& \multicolumn{2}{c|}{0.034}  
& \multicolumn{2}{c|}{0.044} 
& \multicolumn{2}{c}{0.028} 

\end{tabular}
\caption{F-scores and KL-divergences for the system outputs measured against the reference labels. $F_m$ and $F_M$ denote micro F-score and macro F-score, respectively.}
\label{tab:fscores}
\end{table*}

Table \ref{tab:fscores} shows micro  and macro  F-scores of the trained systems and sampling methods. Micro F-score ($F_m$) is the global average over all the data points, whereas macro F-score ($F_M$) is the unweighted average of the class-wise F-scores. The numbers in the table are jackknife estimates of means over ten runs. The confidence intervals are omitted from the table for brevity. In addition to the F-scores, we include the Kullback-Leibler divergences to have another perspective for the system outputs. The results show that the soft metrics correlate well with the hard metrics. However, the soft metrics tend to give higher values since all the common content between the predictions and the references is taken into account. Both soft and hard F-scores agree that DWSNet has the best performance, with the exception of the hard macro F-score, for which the confidence intervals of DWSNet and baseline overlap. Similarly, all the F-scores for TinyCRNN are lower than those of the baseline. This ranking of the trained systems is also in agreement with the KL-divergences. Furthermore, the soft F-scores are also placing the random methods in the same order as the KL-divergence. The OT F-scores for all three random cases are close to each other, therefore it is difficult to draw any conclusion on their difference.

Interestingly, the OT F-score is very similar irrespective of the beta sampling method. The data consists of five different scenes having different sets of classes, whereas the sampling is done independently for each class throughout all the data. As expected, shuffling the order of the distributions takes the predicted labels even further from the references, as shown by the KLD values. The hard and soft F-scores are lower for the shuffled case, but because the optimal threshold follows the distribution of the predictions, the OT F-score fails to indicate that these predictions are worse than the class-wise beta ones.

The confidence intervals of the optimal threshold hard metric and the soft metric are shown in Figure \ref{fig:ci}. The figure indicates that the soft F-scores are more stable than the optimal threshold ones, having narrower CIs, particularly for the micro F-score. The OT method is optimizing threshold values for each class individually, which is causing some variations in the micro F-scores. In addition, there is some overlapping in the confidence intervals of the OT F-scores.

\begin{figure}
    \centering
    \vspace{-15pt}
    \includegraphics[width=65.0mm]{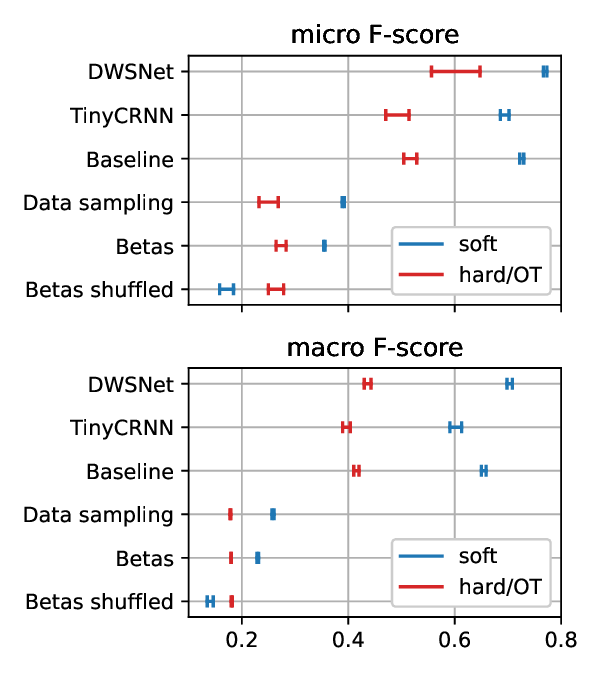}
    \vspace{-10pt}
    \caption{95\% confidence intervals for the average F-score of the trained models and generated random system outputs.}
    \label{fig:ci}
\end{figure}

\subsection{Random system output}

We generate several different random predictions from arbitrarily chosen distributions to investigate the stability of the measures. The predicted labels are sampled from beta distributions with \mbox{$\alpha = \beta = r$} with $r \in \{0.01, 0.1, 1, 5, 20\}$ and evaluated against the challenge reference data. Increasing the parameter $r$ lowers variance and makes the probability mass more concentrated around the middle point. In addition to the random samples we also include a system output of a constant value 0.5. In all the cases, the distribution is symmetric and has expectation of 0.5. 

Due to the symmetry, the hard F-score gives the same result for all the distributions, and is therefore left out from the comparison. Figure \ref{fig:rnd_kld} shows that also the OT hard macro F-score stays constant, but optimizing the class-wise thresholds introduces some variation to the micro F-score. In contrast to the hard F-scores, the soft F-score decreases as the KL-divergence between the predictions and reference labels increases. Despite that the setup is very particular and the generated random samples are far from a real model output, this experiment shows that there are cases where the hard metrics are not able to make the distinction between system outputs with different KL-divergences. 

\begin{figure}
    \centering
    \vspace{-10pt}
    \includegraphics[width=65.0mm]{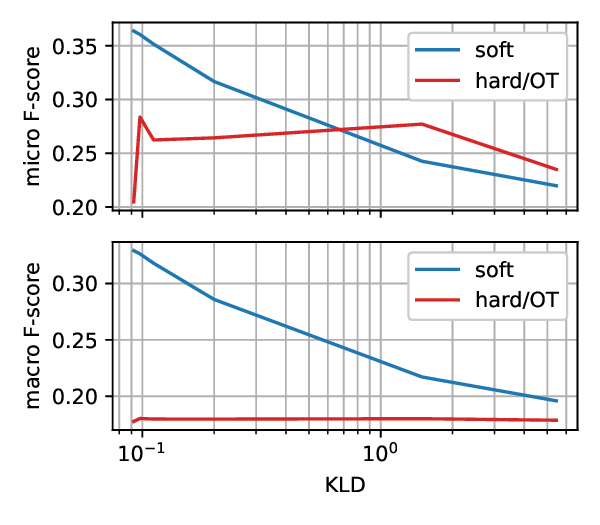}
    \vspace{-10pt}
    \caption{F-score against KL-divergence for the random system predictions evaluated against real data reference labels.}
    \label{fig:rnd_kld}
\end{figure}

\section{Discussion}
\label{sec:discussion}

The hard metrics work well if the reference labels are clearly clustered into positives and negatives. However, when the label values come from a wider scale, converting the data into binary can make the predictions look worse than they are. In practice, the errors from converting the data into binary are not very likely to mislead in the overall interpretation of the results, since averaging over a large number of segments will smooth out the values. However, a non-informative evaluation result is still possible if most of the data lie close to the threshold value. Nevertheless, hard metrics are not useful if we want to measure how well a system learns the uncertainties in the reference labels. 

The soft metrics tend to give higher values than the hard ones since all the segments with nonzero reference or nonzero prediction contribute to the total scores. However, even if the soft F-scores are pushed higher in comparison with the hard ones, the differences between the trained systems are more visible, as the confidence intervals are separated by wider margins.


Finally, the definitions of the soft precision and recall make them differentiable, which makes it possible to train a model using a loss based on the soft metrics. Based on brief experiments with the baseline model, the soft F-score based loss did not bring any significant advantage compared to the MSE loss used in the baseline. Nevertheless, this direction may require more extensive experiments and system architecture design that was out of the scope of this work.

\section{Conclusions}
\label{sec:conclusions}

This paper introduced soft definitions for the precision and recall based on fuzzy sets, extending the classical hard metrics to evaluate system outputs against soft labels. The experiments show that the proposed metrics correlate well with the hard F-scores, and ranking system outputs according to the soft F-score agrees with the ranking based on KL-divergence. Furthermore, the confidence intervals of the results suggest that the soft F-score is more stable than the optimal threshold method.
In future work, we plan to investigate the behavior of the proposed metrics with more real system outputs along with the existing metrics. In addition, the soft precision and recall based loss for model training could be studied in more detail.

\bibliographystyle{IEEEtran}
\balance
\bibliography{refs}
\end{sloppy}
\end{document}